\documentstyle[psfig,prb,twocolumn,aps]{revtex}
\begin{document}
\draft 
\title{Influence of Hybridization on the Properties of the Spinless 
Falicov-Kimball Model}
\author{G.Czycholl}
\address{Department of Physics, University of California San Diego
 La Jolla, CA 92093, USA\\ and\\
Institut f\"ur Theoretische Physik, Universt\"at Bremen, 
          D-28\,334 Bremen, Germany\thanks{permanent address} }
\date{\today} 
\maketitle
\begin{abstract}
Without a hybridization between the localized f- and the conduction (c-) 
electron states the spinless Falicov-Kimball
model is exactly solvable in the limit of high spatial dimension, $ d
\rightarrow \infty$, as first shown by Brandt and Mielsch. 
Here I show that at least for sufficiently small c-f-interaction  
this exact inhomogeneous ground state is also obtained 
in Hartree-Fock approximation. With hybridization the model is no longer
exactly solvable, but the approximation yields that the 
inhomogeneous charge-density wave (CDW) ground
state remains stable also for  finite hybridization $V$  smaller than
 a critical hybridization $V_c$, above which
 no inhomogeneous CDW solution but only a
homogeneous solution is obtained. The spinless FKM does not allow for a 
''ferroelectric'' ground state with a spontaneous polarization, i.e. there
is no 
nonvanishing $<c^{\dagger}f>$-expectation value in the limit of 
vanishing hybridization.
\end{abstract} 
%
%
\section{Introduction}
The spinless Falicov-Kimball model (FKM)\cite{FKM69} without hybridization is
one of the simplest non-trivial interacting many-body models. It takes into
account
a conduction (c-) band, a lattice of localized f-electron states and an on-site
Coulomb (Falicov-Kimball-) interaction between the f- and c-electrons. It was
originally   proposed as a model for (discontinuous) valence and/or
metal-insulator transitions, but has also been interpreted as a model for
crystallization (interpreting the f-''electrons'' as ions or nuclei) 
and for ordering in binary alloys  \cite{JKF93}. The FKM can also be
considered to be a simplified version of the Hubbard model in which the
electrons of the one (spin) species (corresponding to the ''f-electrons'') 
have no dispersion \cite{vanDongVoll}. 

Much of the older work
on the FKM \cite{FKM69,dSF75,AG75} was based on mean-field like
decouplings
of the interaction, and evidence for discontinuous phase (valence)
transitions was obtained. But Leder \cite{Leder78} pointed out that in the
presence of a finite hybridization between f- and c-electrons only
continuous transitions are present in a  Hartree-Fock-decoupling; this
result was confirmed in higher-order decouplings \cite{BaeckCz}. This work
\cite{Leder78,BaeckCz} emphasized, in particular, the importance of a
decoupling 
with respect to excitonic $<c^{\dagger}f>$-expectation values 
in the presence of a finite hybridization. Subsequently it was found by
Brandt and Schmidt \cite{BrSchm86} and, independently, by Kennedy and Lieb
\cite{KennLieb} that for dimension $d \geq 2$ at half filling the FKM 
without hybridization  has an inhomogeneous ground state 
with a chess-board like distribution of the f-particles. Thus a phase
transition from a homogeneous high temperature phase to an inhomogeneous
charge-density wave (CDW) like low temperture phase has to be expected.
Shortly after the discovery of the limit of high dimensions, $ d \rightarrow
\infty$, \cite{MetzVoll} for correlated lattice electron models 
Brandt and Mielsch \cite{BrandtMielsch} found the exact solution of the FKM
without hybridization in this limit. They could obtain exact
results for the band-(c-)electron selfenergy and Green function and for the
f-electron occupation number and confirmed the existece of the phase
transition and of the inhomogeneous CDW- (chess-\-board-) ground state for
large d. To
obtain this solution the FKM has to be mapped on an effective single-site
problem in an auxiliary external field, and this idea is just the foundation
of the nowadays so successful ''dynamical mean-field theory'' (DMFT)
\cite{GeorgesKotliaretal}, which
becomes exact for large-d correlated electron systems. But in contrast to
other correlated electron models (like the Hubbard
model)\cite{GeorgesKotliaretal} one does not rely on approximate or numerical 
(e.g. quantum Monte-Carlo) methods, but the DMFT-equations can 
be solved exactly for the c-electron selfenergy (the ''dynamical mean-field'') 
of the FKM. More recently the  optical properties of the FKM were
investigated by Portengen, \"Ostreich and Sham \cite{POSham96}.  They
studied the FKM with a k-dependent hybridization (taking account of more 
realistic symmetries of the c- and f-states) in Hartree-Fock approximation
and found, in particular, that a non-vanishing excitonic 
$<c^{\dagger}f>$-expectation
value exists even in the limit of vanishing hybridization $V \rightarrow
0$. As an applied (optical) electrical field provides for excitations
between c- and f-states and thus for a polarization expectation value
$P_{cf}=<c^{\dagger}f>$, the finding of a ''spontaneous'' $P_{cf}$ (without
hybridization or electric field) has been interpreted as evidence for
electronic ferroelectricity.\cite{POSham96} Within the Brandt-Mielsch ground
state, however, one has $P_{cf} = 0$. Therefore, the question arises: Is 
the Brandt-Mielsch ground state stable against a small, finite hybridization 
or is one led to a different ground state if one starts from a finite
hybridization and studies the $V \rightarrow 0$ limit of the model? Does
the spinless FKM with an (infinitesmal) small hybridization still have the
inhomogeneous CDW-like ground state \cite{BrandtMielsch} or does it have a
ferroelectric ground state with a spontaneous, non-vanishing polarization 
$P_{cf}$\cite{POSham96}?

These are the questions to be answered in this contribution. To my knowledge it 
has not yet been investigated, which influence a finite hybridization has on 
the
Brandt-Mielsch solution; in fact, with a few exceptions \cite{Hong,POSham96} 
 most recent 
FKM-work\cite{BrandtMielsch,GeorgesKotliaretal,JKF93,vanDongVoll,Fark95}
 neglected the hybridization. Of course, with hybridization an exact solution
of the FKM is no longer possible but one has to rely on approximations. It is
easy to see analytically that in the weak-coupling limit (of small interaction)
the exact Brandt-Mielsch solution contains the Hartree-Fock solution in lowest 
order. The inhomogeneous CDW-Brandt-Mielsch ground state is even exactly
reproduced within the Hartree-Fock treatment of the FKM without hybridization. 
Comparing the CDW-order parameter obtained in the exact Brandt-Mielsch-solution 
and in the Hartree-Fock approximation one sees that they completely agree for 
zero temperature, i.e. for the ground state, but the critical temperature $T_c$
obtained in Hartree-Fock approximation is way too large compared to the 
Brandt-Mielsch result for $T_c$. For low temperatures, however, the 
Hartree-Fock result is reliable, and the Hartree-Fock treatment can, of 
course, also be applied to the FKM with hybridization. Then one obtains an
inhomogeneous CDW-phase at low temperatures also for finite but small 
hybridization $V < V_c$; above the critical hybridization $V_c$ , 
however, 
there is no longer an inhomogeneous CDW-phase, but only a spatially 
homogeneous result is obtained within mean-field theory. Thus the inhomogeneous 
CDW-phase exists also within Hartree-Fock theory  for $V < V_c$, but it has 
been overlooked 
in previous Hartree-Fock treatments \cite{Leder78,POSham96} probably because 
a homogeneous solution was anticipated in this work.  For finite $V$ there 
exists, of course, a finite polarization $P_{cf} = <c^{\dagger}f>$, which is 
strongly enhanced due to the Falicov-Kimball interaction. Studying $P_{cf}$ 
as a functiuon of $V$ one sees that it vanishes for $V \rightarrow 0$, if 
one follows the inhomogeneous CDW-solution. If one follows the 
(additionally existing)
 homogeneous Hartree-Fock solution, however, one obtains a polarization 
$P_{cf} \neq 0$ for $V \rightarrow 0$ in agreement with 
Ref.\onlinecite{POSham96}. Thus a
ferroelectric phase with a non-vanishing $P_{cf}$ for vanishing hybridization 
$V = 0$ is obtained within a Hartree-Fock treatment of the FKM, but, 
unfortunately, this is not the most stable (favorable) Hartree-Fock solution.

Of course, this statement concerns only the spinless FKM, which is not  very
realistic, because any real Fermi system has at least a spin degeneracy. In 
fact, I do not know of any real electronic 
system exhibiting the inhomogeneous ground
state with a CDW (chess-board) pattern obtained for the spinless FKM (without
and with hybridization $V \leq V_c$). Therefore, it has still to be investigated
if more realistic models (including spin and orbital degrees of freedom for
the electronic states) would not allow for different ground states, for instance
the ferroelctric one.\cite{POSham96} Nevertheless, the investigation of the 
spinless FKM with hybridization is of interest at least as a model study because
it allows for a study of the influence of the Falicov-Kimball (interband-)
interaction alone, because of the interpretation of the FKM as a simplified
version of the Hubbard model\cite{vanDongVoll}, and because of the exactly 
solvable limit $V \rightarrow 0$.  

The paper is organized as follows: In section 2 I describe the FKM (with and
without hybridization) and the Brandt-Mielsch solution. Section 3 shows that 
the Hartree-Fock solution is obtained from the exact Brandt-Mielsch solution in
the weak-coupling limit and that the Hartree-Fock approximation yields the
exact ground state properties of the model without hybridization. Numerical
Hartree-Fock results for the FKM with hybridization are presented in Section 4
indicating that also for a small finite hybridization $V \leq V_c$ the
ground state is of the inhomogeneous (chess-board) type, and the final Section 5
contains a short conclusion.

%
%
\section{Falicov-Kimball Model and Brandt-Mielsch Solution}
The spinless Falicov-Kimball model (FKM) is defined by the Hamiltonian

\begin{equation}
H = H_0 + H_1
\label{hamilton}
\end{equation}
Here

\begin{eqnarray}
H_0 &=& \sum_{\bf k} \varepsilon_{\bf k} c_{\bf k}^{\dagger} c_{\bf k} + 
      \sum_{\bf R} E_f f_{\bf R}^{\dagger} f_{\bf R} +
     \sum_{\bf R} U c_{\bf R}^{\dagger} c_{\bf R} f_{\bf R}^{\dagger}f_{\bf R}
 \\
    &=& \sum_{\bf R} \left( \sum_{{\bf \Delta} n.n.} t c_{\bf R}^{\dagger} c_{\bf R
+ \Delta} + E_f f_{\bf R}^{\dagger} f_{\bf R} +
      U c_{\bf R}^{\dagger} c_{\bf R} f_{\bf R}^{\dagger}f_{\bf R} \right)
\nonumber
\label{H_0}
\end{eqnarray}
describes the FKM without hybridization. It consists of a conduction (c-)band
with a tight-binding dispersion

\begin{equation}
\varepsilon_{\bf k} = t \sum_{{\bf \Delta} n.n.} \exp{(i {\bf k \bf \Delta})}
\end{equation}
where $t$ is the nearest neighbor hopping, ${\bf \Delta}$ denotes 
nearest-neighbor (n.n.)
lattice vectors and the band center has been chosen as the zero of the
energy scale, a lattice of f-electron states localized at the lattice sites
${\bf R}$ with   energy $E_f$, and a short ranged Coulomb (Falicov-Kimball-)
interaction $U$ between c- and f-electrons at the same lattice sites. The second
part

\begin{equation}
H_1 = V \left(f_{\bf R}^{\dagger} c_{\bf R} + c_{\bf R}^{\dagger} f_{\bf R}
\right)
\end{equation}
describes the hybridization between the conduction and f-electron states,
which for simplicity is assumed to be also on-site (local) in the present
simple model study, though for
realistic f- and conduction-electron (d- or s-band) states a hybridization
must have a k-dependence (dispersion) for symmetry (parity)
reasons.\cite{POSham96} 

The large-d limit for such a correlated lattice electron model is defined as
the limit $d \rightarrow \infty, t \rightarrow 0$ keeping $d t^2 =
const.$.\cite{MetzVoll} Then for a d-dimensional (hyper)cubic lattice with
nearest neighbor hopping the unperturbed tight-binding density of states of
the conduction electron band is a Gaussian function $\rho_0(E) =
\exp{(-E^2)}/\sqrt{\pi}$ \cite{MetzVoll}. Instead of this a semielliptic model
density of states will be considered here, i.e.

\begin{equation}
\rho_0(E) = \frac{2}{\pi} \sqrt{1-E^2}
\end{equation}
This semielliptic density of states becomes correct for a Bethe lattice in
the limit of a large coordination number, and it may be considered to be
advantegeous compared to the Gaussian density of states, because it has a
finite bandwidth and the correct 3-dimensional bandedge van-Hove
singularities so that it may better model realistic 3-dimensional systems.
The corresponding unperturbed band-electron one-particle Green function is
given by

\begin{equation}
F_0(z) = \frac{1}{N} \sum_{\bf k} \frac{1}{z - \varepsilon_{\bf k}} =
2.\left(z - \sqrt{z^2 - 1}\right)
\end{equation}
Thereby half the unperturbed conduction band width has been chosen as the
energy unit. The explicit form of the unperturbed conduction density of states 
is not important; one could also use a realistic 3-dimensional tight-binding
density of states. The important issue to be adopted from the large-d limit is
the fact that the selfenergy (the ''dynamical mean field'') 
of the interacting system is site-diagonal
(local).\cite{MetzVoll,GeorgesKotliaretal} Therefore the selfenergy
is a functional of the local Green function alone and the functional
dependence is the same as that of an effective atomic or single-impurity
problem.\cite{BrandtMielsch} For the FKM without hybridization ($V=0$), 
i.e. for the
Hamilton $H_0$ alone, which case will only be discussed in the remainder of
this section, Brandt and
Mielsch could analytically determine this functional for the conduction
electron selfenergy, namely\cite{BrandtMielsch}

\begin{equation}
\Sigma_{\bf R} (z) = \frac{U n_{f \bf R}}{1 - (U - \Sigma_{\bf R}(z))
G_{c\bf R}(z)}
\label{selfenfunct}
\end{equation}
Here $G_{c\bf R}(z)$ is the on-site matrix element of the full conduction
electron Green function $G_c(z)$, which depends itself on the selfenergy to
be determined, and $n_{f\bf R}$ is the f-electron occupation number at site
$\bf R$. The selfenergy functional (\ref{selfenfunct}) is just of the
Hubbard-III (alloy-analog) form, or, in other words,
 it is just the selfenergy functional of
the coherent potential approximation (CPA) for disordered alloys, if the
f-electron occupation number $n_{f\bf R}$ is interpreted as an impurity
concentration. But here the f-electron occupation of the sites does not
occur at random, but it depends itself on the band-electron Green function
and the occupation of the other sites. Explicitly the f-electron occupation
number is given by \cite{BrandtMielsch,JKF93}

\begin{equation}
n_{f \bf R} = \frac{1}{1 + \exp{\left((E_f-\mu)/T\right)}
\prod_n \left(1-UG_{c{\bf R}0}(z_n)
\right)^{-1}}
\end{equation}
where, as usual, $T$ is the temperature (measured in energy units, i.e. 
$k_B = 1$), $\mu$ is the chemical potential, $z_n = \mu + i \omega_n =
\mu + i (2n+1)\pi T$ denotes the Matsubara frequencies  and

\begin{equation}
G_{c{\bf R}0}(z) = \frac{G_{c{\bf R}}(z)}{1 + \Sigma_{\bf R}(z) G_{c \bf R}(z)}
\end{equation}

Obviously the f-electron occupation number can also be written in the form
\cite{SiKotliar92}

\begin{equation}
n_{f \bf R} = f(\tilde{E}_{f \bf R})
\end{equation}
where $f(E) = [\exp{((E - \mu)/T)} + 1]^{-1}$ denotes the Fermi function
and the effective f-level energy $\tilde{E}_{f \bf R}$ is given by

\begin{equation}
\tilde{E}_{f \bf R} = E_f - T \sum_n \ln{(1 - U G_{c {\bf R}0}(z_n))} 
\label{efff-lev}
\end{equation}

In the remainder of the paper only the case of half filling, i.e. one electron
per lattice site $n_{f \bf R} + n_{c \bf R} = 1$, and the symmetric model, i.e.
$E_f = 0$ will be considered; then the chemical potential is automatically 
fixed at  $\mu = U/2$. 

If one anticipates a translationally invariant form of the solution, 
the selfenergy $\Sigma_{\bf R}(z)$, band-electron Green function 
$G_{c \bf R}(z)$ and f-electron occupation number $n_{f \bf R}$ are 
translationally invariant and do not depend on the lattice site ${\bf R}$. Then
the Green function is simply given by

\begin{equation}
G_{c \bf R}(z) = G_c(z) = \frac{1}{N} \sum_{\bf k} 
\frac{1}{z - \Sigma(z) - \varepsilon_{\bf k}} = F_0(z-\Sigma(z))
\end{equation}
But in general an inhomogeneous solution has to be expected for the ground state,
as the following simple consideration for the atomic limit (i.e. hopping $t=0$)
 shows:  If one assumes that in the atomic limit the average occupation number
for  both, the f- and c-electrons is $1/2$, then, 
of course, a spatial separation of
f- and c-electrons is energetically most favorable, i.e. half of the lattice
sites is filled with f- and the other half with c-electrons; then because
of the short-ranged nature of the model interaction all interactions 
(repulsions) are avoided and all electrons have only their one-particle
energies $E_f = E_c$  ($=0$ for the above choice of the energy scale). 
Because of the competition with the kinetic energy, 
the picture is no longer as simple in the case of a finite conduction 
band width (finite hopping $t$).
Then also the sites occupied by f-electrons get a finite occupation 
probability for band-electrons. But nevertheless an inhomogeneous ground state 
is energetically most favorable. For a bipartite lattice with an  A- and
B-sublattice in the ground state  the f-electron states are
occupied for the sites of one sublattice, say A, and empty for ${\bf R}$
$\epsilon$ $B$, and therefore the band electrons see an alternating effective
potential and will also inhomogeneously (alternating from A- to B-site)
be occupied. Defining the A-(B-) occupation numbers as

\begin{eqnarray}
n_{aA}(T) = n_{a {\bf R}} = <a_{\bf R}^{\dagger}a_{\bf R}> \mbox{ for {\bf R} }
\epsilon \mbox{ } A  \nonumber \\
n_{aB}(T) = n_{a {\bf R}} =  <a_{\bf R}^{\dagger}a_{\bf R}> \mbox{ for {\bf
R} } \epsilon \mbox{ } B 
\end{eqnarray}
and $ a \epsilon \{c,f\}$ one has $n_{fA}(T=0) = 1$, $n_{fB}(T=0) = 0$ and 
$0 \leq n_{cA}(T=0) < n_{cB}(T=0) \leq 1$ and can define the CDW order
parameter as

\begin{equation}
m(T) = n_{cB}(T) - n_{cA}(T)
\end{equation}

A full polarization of the band electrons, i.e. $m(T=0) \rightarrow 1$, can be
expected only in the strong coupling (large $U$) limit. It has been shown
\cite{BrSchm86,KennLieb,BrandtMielsch} that for 
half filling the ground state of the FKM without hybridization has, in fact,
the chess-board symmetry, i.e. a finite CDW-order parameter $m(T=0)$.
Furthermore, for $d \rightarrow \infty$ correlation functions, the critical
temperature $T_c$ at which the order parameter vanishes (as a function of
the correlation U), the free energy and other quantities  could be
calculated \cite{BrandtMielsch}; away from half filling indications for
phase separation were obtained.\cite{JKF93,BrandtMielsch}

\section{Brandt-Mielsch and Hartree-Fock Solution for Vanishing
Hybridization}

In this section I will show that for small interaction $U < 1$ the
Hartree-Fock approximation
 becomes reliable and contains the most essential features of
the exact Brandt-Mielsch solution (partially even quantitatively) for the
FKM without hybridization. From the exact results
(\ref{selfenfunct}) and (\ref{efff-lev}) one obtains in lowest order in the
correlation for the band-electron selfenergy

\begin{equation}
\Sigma_{\bf R}(z) = U n_{f \bf R}
\label{HartrFock1}
\end{equation}
and for the effective f-level energy

\begin{equation}
\tilde{E}_{f \bf R} = E_f + U T \sum_n G_{c \bf R}(z_n) = E_f + U <n_{c \bf
R}>\label{HartrFock2}
\end{equation}
Obviously this is just the standard Hartree-Fock approximation for the FKM
without hybridization, which thus follows from the exact result by an
expansion with respect to $U$ in lowest (linear) order in $U$. This result
is certainly not surprising but rather as it should be expected.

\begin{figure}[bht]
\begin{center}
\parbox{9cm}{
\psfig{file=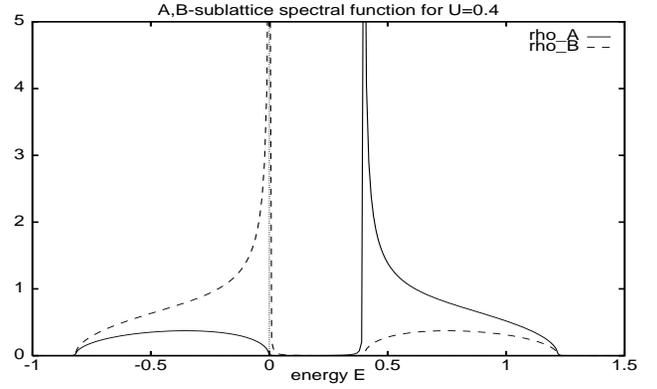,width=9cm,height=5cm,angle=-90}
}
\end{center}

\caption[]{A- and B-sublattice band electron spectral function for 
$U = 0.4, T=0$ in the inhomogeneous CDW phase}
 \label{Fig1}
\end{figure}

For the inhomogeneous CDW-phase the agreement between Hartree-Fock and
exact solution is even stronger and -- at least for $T = 0$, i.e. for the
ground state-- rigorously fulfilled (i.e. without $U$-expansion). Because of
the full polarization of the f-electrons at $T = 0$, i.e. $n_{fA}(T=0) = 1$
and $n_{fB}(T=0) = 0$, one gets from (\ref{selfenfunct}) 

\begin{equation}
\Sigma_{A}(z) = U \mbox{\hspace{1cm}} \Sigma_{B}(z) = 0
\end{equation}
which obviously corresponds to the Hartree-Fock result in this case.
Therefore, if the f-electron system is fully polarized, the band-electron
spectral function obtained in Hartree-Fock approximation 
(shown in Fig. \ref{Fig1}
for U=0.4) is identical to the exact spectral function. But even for 

\vspace{0.2cm}

\begin{figure}[bht]
\begin{center}
\parbox{9cm}{
\psfig{file=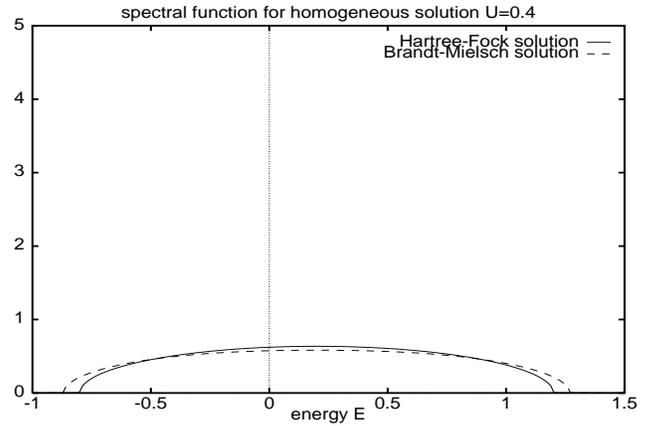,width=9cm,height=5.5cm,angle=-90}}
\end{center}

\caption[]{Band electron spectral function for the homgeneous phase within 
the exact Brandt-Mielsch solution and in Hartree-Fock approximation 
for $U = 0.4$}
\label{Fig2}
\end{figure}

\vspace{0.2cm}

\begin{figure}[bht]
\begin{center}
\parbox{9cm}{
\psfig{file=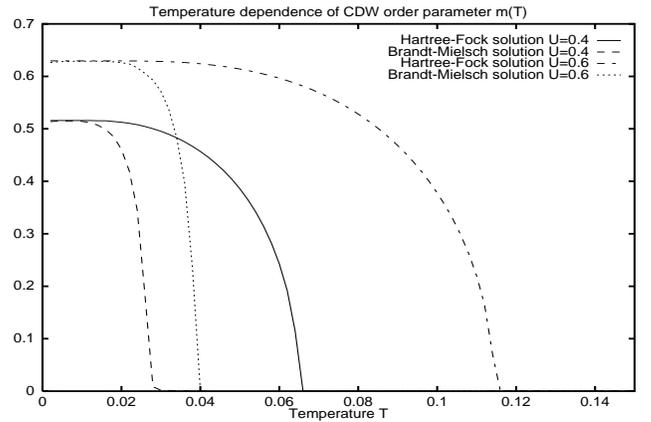,width=9cm,height=5.5cm,angle=-90}}
\end{center}

\caption[]{Temperature dependence of the CDW order parameter $m(T)$ for the 
FKM without hybridization ($V=0$) within the exact Brandt-Mielsch solution and 
within the Hartree-Fock approximation for $U=0.4$ and $U=0.6$}
\label{Fig3}
\end{figure}

\vspace{0.5cm}
\noindent
the
homogeneous phase ($n_{fA} = n_{fB} = 0.5$), which may be artificially
enforced for low $T$ or becomes the only solution for high $T$, the
difference between the Hartree-Fock and the exact solution for the spectral
function may be small, as shown in Fig. \ref{Fig2}; 
this holds true if the correlation
$U$ is sufficiently small so that the (around $\mu = U/2$ symmetric)
exact conduction electron density of states does not show indications of the
splitting into the upper and lower Hubbard band. In Fig. \ref{Fig3}
 I present results
for the temperature dependence of the CDW order-parameter obtained in
Hartree-Fock approximation and exactly for $U=0.4$ and $U=0.6$. Obviously
there is full agreement between the exact and Hartree-Fock result for $m(T)$
for low temperatures $T \rightarrow 0$ in agreement with the above
argumentation. But with increasing temperature (when the f-electrons are no
longer fully polarized) $m(T)$ in Hartree-Fock becomes different from the
exact result and, in particular, the critical temperature $T_c$ above which
the CDW order parameter vanishes is way too large in Hartree-Fock
approximation compared to the exact (Brandt-Mielsch-) result. But at least for 
the low temperature (ground state) properties and small to intermediate values 
of the cf-correlation $U$ ($U \leq 1$), which is realistic for the 
Falicov-Kimball interaction, the Hartree-Fock approximation yields already 
good results for the FKM without hybridization in agreement with the exact 
Brandt-Mielsch result.

%
%
\section{ Hartree-Fock Treatment for Finite Hybridization} 

The FKM with hybridization is no longer exactly solvable, but, of course, 
approximations like the Hartree-Fock-treatment can be applied. As these 
Hartree-Fock results become correct for the groundstate properties of the
model without hybridization, it can be expected that this simple
approximation also reproduces the essential ground state properties of the
FKM with hybridization, at least qualitatively and for $U \leq 1$. Within
Hartree-Fock theory the FKM Hamiltonian  (\ref{hamilton}) is replaced by the
following effective one-particle Hamiltonian:

\begin{eqnarray}
H_{eff} = \sum_{\bf R} && \left( \tilde{E}_{c \bf R} c_{\bf R}^{\dagger} 
c_{\bf R} + t
\sum_{{\bf \Delta} n.n.} c_{\bf R}^{\dagger} c_{\bf R + \Delta} 
 +  \tilde{E}_{f \bf R} f_{\bf R}^{\dagger} f_{\bf R} \right. \nonumber \\
 &+& \left.  
\tilde{V}_{\bf R} (f_{\bf R}^{\dagger} c_{\bf R} + c_{\bf R}^{\dagger}
f_{\bf R}) \right)
\label{HartFock}
\end{eqnarray}
where the effective parameters are given by

\begin{eqnarray}
\tilde{E}_{c \bf R} = U n_{f \bf R} \nonumber \\
\tilde{E}_{f \bf R} = E_f + U n_{c \bf R} \nonumber \\
\tilde{V}_{\bf R} = V - U P_{cf \bf R} =  V - U 
<c_{\bf R}^{\dagger}f_{\bf R}>
\label{HFpar} 
\end{eqnarray}
and have to be determined selfconsistently. Of course, as the exact solution 
for $V = 0$ is of the spatially inhomogeneous CDW type, one has to allow for an
inhomogeneous Hartree-Fock solution in the case of finite hybridization $V$, 
too, i.e. one should not only look for a translationally invariant 
 solution of (\ref{HartFock}), 
for which the expectation values $n_f, n_c, P_{cf}$ 
are independent on the lattice site ${\bf R}$, as it has been done in 
previous Hartree-Fock studies of the FKM with hybridization.
\cite{Leder78,POSham96} 
Again different expectation values and thus effective one-particle parameters
will be admitted for lattice sites from the A- or B-sublattice. 
The expectation values are given by

\begin{eqnarray}
n_{c \bf R} = <c_{\bf R}^{\dagger}c_{\bf R}> = -\frac{1}{\pi} \int dE
f(E) \mbox{Im} G_{c \bf R}(E+i0) \nonumber \\
n_{f \bf R} = <f_{\bf R}^{\dagger}f_{\bf R}> = -\frac{1}{\pi} \int dE
f(E) \mbox{Im} G_{f \bf R}(E+i0) \nonumber \\
P_{cf \bf R} = <c_{\bf R}^{\dagger}f_{\bf R}> = -\frac{1}{\pi} \int dE
f(E) \mbox{Im} G_{fc \bf R}(E+i0)
\end{eqnarray}
In the case of a finite hybridization and  a possible AB-sublattice 
structure the
on-site matrix elements of the Green function are explicitely given by

\begin{eqnarray}
G_{f \bf R}(z) = \frac{1}{z - \tilde{E}_{f \bf R}} \left(1 + 
\tilde{V}_{\bf R}
G_{fc \bf R}(z) \right) \nonumber \\
G_{fc \bf R}(z) = \frac{\tilde{V}_{\bf R}}{z - \tilde{E}_{\bf R}} G_{c \bf R}(z)
\nonumber \\ 
G_{c A(B)}(z) = \sqrt{\frac{Z_{B(A)}}{Z_{A(B)}}}F_0(\sqrt{Z_AZ_B})
\end{eqnarray}
with

\begin{equation}
Z_{A(B)} = Z_{\bf R} = z - \tilde{E}_{c \bf R} - \frac{\tilde{V}_{\bf R}^2}
{z - \tilde{E}_{f \bf R}} \mbox{for  {\bf R} } \epsilon \mbox{ } A 
(\mbox{ \bf R } \epsilon \mbox{ } B)
\end{equation}

The above selfconsistency equations can easily be solved numerically by
iteration, and
some of the results are shown in the following figures. Fig. \ref{Fig4}
shows the temperature dependence of the CDW order parameter $m(T) =
n_{cB}(T) - n_{cA}(T)$ for different values of the hybridization $V$.
According to the discussion in the previous section the $T \rightarrow 0$ 
value of $m(T)$ can be expected to correspond to the exact value, whereas
the critical temperature $T_c$ is probably too large, as in most
Hartree-Fock-studies. Obviously the order parameter strongly decreases with
increasing hybridization. This results from the fact that for any
non-vanishing hybridization the f-electron occupation is no longer a good
quantum number for the FKM; consequently the f-electron occupation number 
of one lattice site is no longer exactly $0$ or $1$ in the ground state, and
therefore the band-electron polarization is also smaller than without
hybridization and is the smaller the larger the hybridization is. Fig.
\ref{Fig5} shows the low temperature order parameter $m(T,V))$ (for $T=0.008$) 
as a function of the hybridization, which confirms the decrease of the
CDW-phase with increasing $V$. Obviously there is a critical value $V_c$ of
the hybridization at which the CDW-order-parameter vanishes. For 

\begin{figure}[bht]
\begin{center}
\parbox{9cm}{
\psfig{file=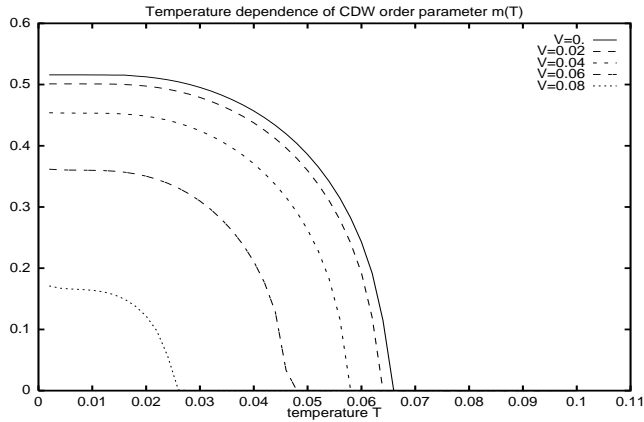,width=9cm,height=5.5cm,angle=-90}}
\end{center}

\caption{Temperature dependence of the Hartree-Fock CDW order parameter $m(T)$
for different hybridizations $V$ and $U=0.4$}
\label{Fig4}
\end{figure}

\vspace{0.5cm}

\begin{figure}[bht]
\begin{center}
\parbox{9cm}{
\psfig{file=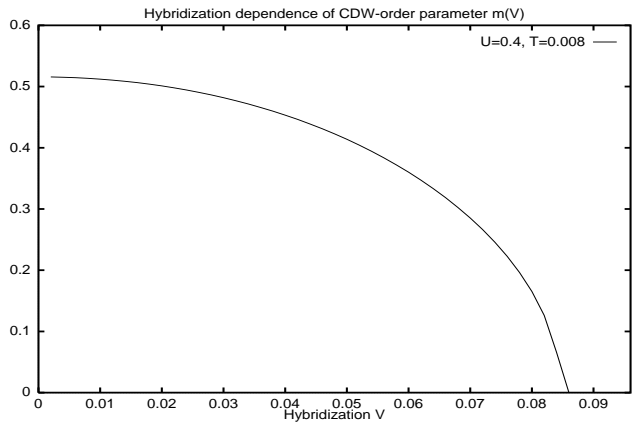,width=9cm,height=5.5cm,angle=-90}}
\end{center}

\caption{Hybridization dependence of the Hartree-Fock CDW order parameter $m(V)$
for $U=0.4$, at low temperature ($T=0.008$)}
\label{Fig5}
\end{figure}

\begin{figure}[bht]
\begin{center}
\parbox{9cm}{
\psfig{file=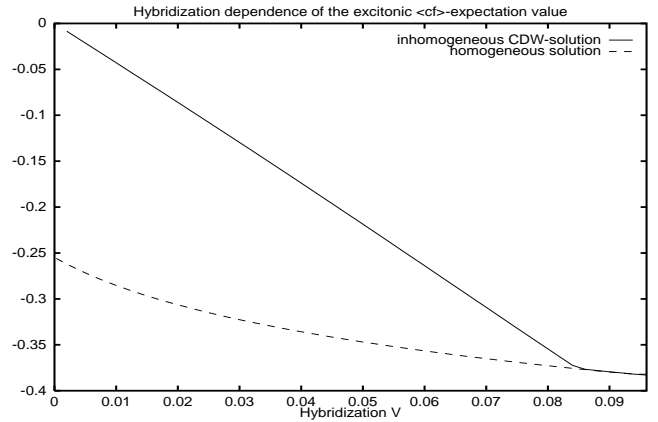,width=9cm,height=5.5cm,angle=-90}}
\end{center}

\caption{Hybridization dependence of the c-f-polarization 
$P_{cf} = <c_{\bf R}^
{\dagger}f_{\bf R}>$ for the inhomogeneous CDW-solution (full line) and for
the translationally invariant homogeneous solution (dashed line)}
\label{Fig6}
\end{figure}

\vspace{0.5cm}

\noindent
larger $V$
only a homogeneous, translationally invariant solution exists. Finally the
hybridization dependence of the off-diagonal expectation value $P_{cf \bf R} = 
<c_{\bf R}^{\dagger}f_{\bf R}> $ is shown in Fig. \ref{Fig6}. Obviously, for
the most stable, inhomogeneous CDW-solution $P_{cf \bf R}$ vanishes linearly
with the hybridization:

\begin{equation}
P_{cf \bf R} = <c_{\bf R}^{\dagger}f_{\bf R}> \rightarrow 0 \mbox{ for } V
\rightarrow 0
\end{equation}

But the above Hartree-Fock selfconsistency equations have also a
homogeneous, translationally invariant solution, for which the CDW order
parameter $m(T)$ vanishes and for which  $P_{cf}(V)$ is given
by the dashed line in Fig. \ref{Fig6}, if one starts from the
high-V  homogeneous solution and follows it to small values of V. Obviously,
for the homogeneous Hartree-Fock solution one has $P_{cf}(V) \neq 0$ for $V
\rightarrow 0$. Then one would have a built-in polarization,\cite{POSham96},
 a nonvanishing
excitonic expectation value $<c_{\bf R}^{\dagger}f_{\bf R}>$ for vanishing
hybridization, a different kind of symmetry breaking formally resembling
superconductivity (also concerning the type of selfconsistency equation and
of the temperature dependence of the order parameter $P_{cf}(V=0,T)$). But
this homogeneous Hartree-Fock solution is not the most stable one, the
inhomogeneous CDW solution has the lower energy as becomes clear
immediately from the densities of states of Fig. \ref{Fig1},\ref{Fig2}.

\vspace{0.5cm}

\vspace{0.5cm}

%
\section{Summary and Conclusion} 
To summarize, I have studied the simple spinless Falicov-Kimball model for
half filling (i.e. one elctron per lattice site) and in the symmetric case
and found that without hybridization the simple Hartree-Fock approximation
reproduces the exact CDW ground state first obtained by Brandt and coworkers
\cite{BrSchm86,BrandtMielsch}. Furthermore I found that also for finite but
small hybridization, for which no exact result is available, an 
inhomogeneous CDW
ground state is obtained, i.e. this ground state remains stable when turning
on a hybridization (or applying a field providing for c-f-transitions and
thus for an effective hybridization). But the CDW-order parameter decreases
with increasing hybridization, and a critical value of the hybridization
$V_c$ exists above which only a homogeneous solution and no CDW phase is
obtained. For the inhomogeneous phase the ''polarization'' (or the 
c-f-transition rate $<c_{\bf R}^{\dagger}f_{\bf R}>$) vanishes in the limit
$V \rightarrow 0$; so there is no built-in polarization, which the  for
$V \rightarrow 0$ non-vanishing
$P_{cf}$ of the (unstable) homogeneous Hartree-Fock solution might suggest.

Nevertheless, the principal idea of Ref. \onlinecite{POSham96} remains valid, 
namely that the Falicov-Kimball interaction may be of importance 
in particular for
a description of the optical properties of correlated electron systems. When 
there is a (small) finite hybridization (or field providing for c-f-transitions)
the effective hybridization $\tilde{V}$ (cf. Eq. \ref{HFpar})
 is strongly enhanced due to the Falicov correlation. Furthermore,  if the 
one-particle hybridization has a dispersion (e.g. the for parity reasons more
realistic p-wave symmetry), the Falicov-Kimball interaction provides for an
effective hybridization, which has also an on-site (s-wave) component, and this
may be of importance for an understanding of the gap formation in heavy-fermion
(Kondo-) insulators. Finally, it is certainly not excluded that models with
more realistic (spin and orbital) degeneracy  still have the built-in 
polarization, which has been interpreted as electronic ferroelectricity.

\vspace{1cm}

{\bf Acknowledgement}

I thank Lu Sham for his hospitality at UCSD, for suggesting this investigation
to me and for useful discussions.
This work has been supported by the Deutsche Forschungsgemeinschaft (project 
Cz 31/11-1) and by the NSF grant DMR-9721444.

%
%

\end{document}